\documentclass[preprint,aps,pra,superscriptaddress]{revtex4}
\usepackage{tabularx}
\usepackage{dcolumn}
\usepackage{graphics}
\usepackage{bm}
\usepackage[dvips]{graphicx}
\usepackage{tabularx}
\usepackage{amsmath}
\usepackage{amssymb}
\usepackage{longtable}
\usepackage{verbatim}
\usepackage{float} 
\usepackage{fancyhdr} 
\usepackage{amsmath}
\usepackage{multirow}
\usepackage{comment}

\begin{document}
\pagestyle{fancy}
\rhead{\thepage}

\title{Charmonium spectrum in an unquenched quark model}
\author{Sadia Kanwal}
\affiliation{Centre For High Energy Physics, University of the Punjab, Lahore(54590), Pakistan. }
\author{Faisal Akram}
\affiliation{Centre For High Energy Physics, University of the Punjab, Lahore(54590), Pakistan. }
\author{Bilal Masud}
\affiliation{Centre For High Energy Physics, University of the Punjab, Lahore(54590), Pakistan. }
\author{E.S. Swanson}
\affiliation{Department of Physics and Astronomy, University of Pittsburgh, Pittsburgh PA 15260, USA.}

\begin{abstract}
The effects of virtual light quark pairs on the charmonium spectrum are studied. Pair creation is modelled with a ``$^{3}P_{0}$" vertex and intermediate states are summed up to 2S excitations. Quark model parameters are obtained by fitting to 12 well-known charmonium states, allowing for feedback between the decaying particle and the induced mass shifts.  Both of these technical steps are new and improve  agreement with the experimental spectrum. In general, the masses receive small shifts once model parameters are refit. This is true in almost cases except the $\chi_{cJ}(2P)$ multiplet, which experiences upwards mass shifts of order 150 MeV, has the ordering of the multiplet rearranged, and pushes the erstwhile $c\bar{c}$ ${2}^3P_1$ state well above $D^*\bar{D}$ threshold--observations that clarify the nature of the enigmatic $X(3872)$.
\end{abstract}

\maketitle
\section{Introduction}

The quark model has a long and distinguished history as an organizing and predictive tool for hadronic physics\cite{qm-history}. The advent of QCD and lattice field techniques have served to solidify its foundations\cite{qm-qcd}, and its modern application to heavy quark systems has achieved a high level of sophistication and accuracy\cite{GI,hcharmonia,qm-heavy}.

In spite of these successes, a number of well-known deficiencies exist in typical quark models. Experimentally, these include the poor descriptions of heavy quark states such as the $D_{s0}(2317)$, $D_{s1}(2460)$, $Y(4260)$, $X(3872)$, $T_{cc}(3875)$, $P_c(4440)$, and a host of other ``exotics"\cite{review}. These states have engendered much effort in modelling them, with a variety of explanatory mechanisms being proposed. Among these are
weakly bound hadronic ``molecular" states\cite{molecules}, cusp singularities\cite{cusp}, hadrocharmonium\cite{hc}, and diquarks\cite{dq}. On the theoretical side, it is anticipated that the suppression (or, rather, absorption) of the effects of light quarks and of gluons must break down somewhere in the spectrum. Coupling to light quarks induces mixing with the heavy-light continuum and can have important effects near decay thresholds to such channels. This effect is thus of particular relevance to states such as the $X(3872)$, $Z_c(2010)$, $P_c(4440)$, and many others.

Here we address the impact of light quarks on the charmonium spectrum in the context of a simple model of quark pair creation (called the ``$^{3}P_{0}$" model). Incorporating this term in the dynamics permits the coupling of a quark-antiquark Fock state to the meson-meson continuum, which has the effect of ``unquenching" the model (borrowing a term from lattice field theory). Work in this area is as old as the quark model itself, starting with the statement of the Oakes-Yang problem in 1963\cite{OY} (namely, why does the Gell-Mann--Okubo mass formula work in view of continuum thresholds?). We cannot summarise the very extensive subsequent development of the field here. For those interested in learning more we recommend perusing the pioneering work of the Cornell group\cite{cornell},
early work in the light spectrum with a focus on the light scalar mesons\cite{scalars} and baryons\cite{baryons}, later work on the heavy quark spectrum\cite{heavy,K1,14feretti}, and the examination of the phenomenology and theoretical underpinnings of the problem given in Refs.~\cite{GI1,thy,08hloop}.

In section II we elaborate the formalism used to calculate mass shifts through hadron loops. Section III contains our results for mass shifts and the charmonium spectrum after performing a global fit, analysis of the results and a comparison to other models.
 Finally, we summarize our results and comment on the prospects for improving the formalism in Section IV.

\section{Quenched and Unquenched Charmonium}

Our goal is to describe the charmonium spectrum with a nonrelativistic quark model of the type used in Ref. \cite{hcharmonia} (referred to as BGS hereafter) with additional coordinate smearing as employed by Godfrey and Isgur(GI)\cite{GI}. This model is then ``unquenched" with the $^{3}P_{0}$ model and refit to the spectrum, as described below.

\subsection{The Quenched Quark Model for a Charmonium System}

The quark model examined here employs nonrelativistic quark kinematics with a Cornell-type central  interaction and spin-dependent interactions as motivated by one gluon exchange\cite{hcharmonia}.

 \begin{equation}\label{qhamiltonian}
    \hat{H}_{0} = 2{m}_c + \frac{\hat{p}^{2}}{m_{c}} + \mathcal{C}  - \frac{4}{3}\frac{\alpha_{s}(r)}{r} + b r + V_{hyp}(r) + V_{LS} + V_{T},
 \end{equation}
where $\alpha_{s}$ is a running coupling and $b$ is the string tension.  The spin-dependent interaction is described in terms of a hyperfine potential

\begin{equation}\label{potspin}
  V_{hyp}(r) =  \frac{1}{m_{c}^2}\frac{32\pi\alpha_{s}(r)}{9} \tilde{\delta}(r)\, \mathbf{S}_{c}\cdot\mathbf{S}_{\overline{c}},
\end{equation}
with $\mathbf{S}_{c}$ and $\mathbf{S}_{\overline{c}}$ being the spin of charm quark and anti-charm quarks respectively and $\tilde{\delta}(r)=(\sigma_h/\sqrt{\pi})^{3} e^{-\sigma_h^{2}r^{2}}$.
The spin-dependent interaction also contains
spin-orbit coupling and tensor terms described by

\begin{equation}\label{potl.s}
V_{LS}(r)=\frac{1}{m_{c}^2}\left(\frac{2\alpha_{s}(r)}{r^{3}}-\frac{b}{2r}\right)\mathbf{L}\cdot\mathbf{S},
\end{equation}

\begin{equation}\label{potT}
 V_{T}(r) = \frac{4}{m_{c}^2}\frac{\alpha_{s}(r)}{r^{3}}  \, [\mathbf{S}_{c}\cdot \hat{\mathbf{r}}  \, \mathbf{S}_{\overline{c}}\cdot \hat{\mathbf{r}} - \frac{1}{3}  \mathbf{S}_{c}\cdot\mathbf{S}_{\overline{c}}].
\end{equation}

The running coupling follows that of Ref. \cite{GI}:

\begin{equation}\label{runnalpha}
 \alpha_{s}(r)= \sum^{3}_{i=1} \alpha_{i} \frac{2}{\sqrt{\pi}}\int^{\gamma_{i} r}_{0} e^{-x^{2}} dx,
\end{equation}
where $\gamma^{2}_{1}=1/4$, $\gamma^{2}_{2}=10/4$ and $\gamma^{2}_{3}=1000/4$. The spatial dependence has been chosen to reproduce the perturbative behaviour of the QCD running coupling at large momentum. We have modified the original approach by fixing $\alpha_2$ and $\alpha_3$ as linear functions of $\alpha_1$ (such that the perturbative behaviour remains). This allows us to vary $\alpha_0 \equiv \alpha_1+\alpha_2+\alpha_3$ when fitting the spectrum.

%

The delta function that appears in the hyperfine interaction and the
$1/r^{3}$ terms are illegal operators in three-dimensions and hence must be regulated. This is done by smearing according to Ref. \cite{GI}

\begin{equation}\label{smear}
\tilde{f}(r)\;=\;\int d^{3}r'\; f(r') \frac{\sigma_{S}^3}{\pi^{3/2}} e^{-\sigma^{2}_{S}(r-r')^{2}},
\end{equation}
where $\sigma_S$ is another fit parameter.

We remark that this model differs from that of BGS in using a running coupling, smeared coordinates, and evaluating the spin-dependent operators nonperturbatively; it differs from GI in the use of nonrelativistic kinetic energy and neglect of the isoscalar annihilation interaction and  ``relativizing" factors of energy divided by mass.

The model is solved with the shooting method.
Fitting to twelve  well-established charmonia gives the parameters reported in Table \ref{parameters}, and the masses reported in Table \ref{qspectrum}. We note that the constant term is quite small, in keeping with the simpler BGS model. Charmonium masses are similar between the models, with an average deviation of 14 MeV for the BGS model and 17 MeV for the quenched model presented here.

\begin{table}[H]
\caption{Experimental and quenched quark model spectrum of $c\bar{c}$ states in MeV.}
\centering
\footnotesize
\begin{tabular}{cccc|cccc}
  \hline\hline
   Meson state   & Exp. Mass~\cite{pdg} & $M$& BGS\cite{hcharmonia} & Meson state  & Exp. Mass~\cite{pdg} & $M$ & BGS\cite{hcharmonia} \\
  \hline
  $J/\psi (1 ^3 S_{1}) $ & 3096.9 $\pm$ 0.006  & 3090 & 3090 & $\chi_{c2}(5 ^3P_{2})$&           & 4890 &     \\
  $\eta_{c}(1 ^1S_{0})$  & 2983.9 $\pm$ 0.5    & 2983 & 2982 &$\chi_{c1}(5 ^3P_{1})$&           & 4884 &     \\
  $\psi' (2 ^3S_{1})$    & 3686.097 $\pm$ 0.025& 3688 & 3672 &$\chi_{c0}(5 ^3P_{0})$&           & 4869 &    \\
 $\eta^{'}_{c}(2 ^1S_{0})$& 3637.6 $\pm$ 1.2    & 3648 & 3630 &$h_{c}(5 ^1P_{1})$    &           & 4876 &      \\
  $\psi$ $(3^3S_{1})$   & 4039 $\pm$ 1        & 4091 & 4072 &$\psi_{3} (1 ^3D_{3})$  &         & 3847 & 3806\\
  $\psi$ $(4^3S_{1})$   & 4421 $\pm$ 4        & 4426 & 4406 & $\psi_{2} (1 ^3D_{2})$&          & 3803 & 3800\\
  $\chi_{c2} (1 ^3P_{2})$& 3556.17 $\pm$ 0.07  & 3555 & 3556 &$\eta_{c2}(1 ^1D_{2})$    &        & 3812 & 3799\\
  $\chi_{c1} (1 ^3P_{1})$& 3510.67 $\pm$ 0.05  & 3513 & 3505 &$\psi_{3} (2 ^3D_{3})$    &       & 4193 & 4167 \\
  $\chi_{c0} (1 ^3P_{0})$& 3414.71 $\pm$ 0.3   & 3442 & 3424 &$\psi_{2} (2 ^3D_{2})$   &        & 4168 & 4158 \\
  $h_{c}(1 ^1P_{1})$    & 3525.38 $\pm$ 0.11  & 3518 & 3516 &$\eta_{c2}(2 ^1D_{2})$   &        & 4172 & 4158 \\
  $\psi (1 ^3D_{1})$ & 3773.13$\pm$0.35    &3743  & 3785 &$\psi_{3} (3 ^3D_{3})$  &         & 4502 & \\
  $\psi (2 ^3D_{1})$& 4191 $\pm$ 5        & 4135 & 4142 &$\psi_{2} (3 ^3D_{2})$   &        & 4485 & \\
  $\eta_{c}(3 ^1S_{0})$ &                     & 4064 & 4043 &$\psi (3 ^3D_{1})$       &        & 4461 & \\
  $\eta_{c}(4 ^1S_{0})$  &                     & 4404 & 4384 &$\eta_{c2}(3 ^1D_{2})$    &        & 4486 & \\
  $\psi$ $(5^3S_{1})$    &                     & 4723 &      &$\psi_{3} (4 ^3D_{3})$  &         & 4784 &  \\
  $\eta_{c}(5 ^1S_{0})$  &                     & 4705 &      &$\psi_{2} (4 ^3D_{2})$   &        & 4771 & \\
  $\chi_{c2} (2 ^3P_{2})$&                     & 3967 & 3972 &$\psi (4 ^3D_{1})$       &        & 4753 &  \\
  $\chi_{c1} (2 ^3P_{1})$ &                     & 3947 & 3925 &$\eta_{c2}(4 ^1D_{2})$    &        & 4771 &  \\
  $\chi_{c0} (2 ^3P_{0})$ &                     & 3909 & 3852 &$\psi_{3} (5 ^3D_{3})$   &        & 5046 &  \\
  $h_{c}(2 ^1P_{1})$    &                     & 3943 & 3934 &$\psi_{2} (5 ^3D_{2})$   &        & 5036 &   \\
  $\chi_{c2} (3 ^3P_{2})$ &                     & 4309 & 4317 &$\psi (5 ^3D_{1})$       &        & 5020 &    \\
  $\chi_{c1} (3^3P_{1})$ &                     & 4298 & 4271 & $\eta_{c2}(5 ^1D_{2})$    &        & 5035 & \\
  $\chi_{c0} (3 ^3P_{0})$&                     & 4273 & 4202 & & & &\\
  $h_{c}(3 ^1P_{1})$    &                     & 4291 & 4279 & & & &\\
  $\chi_{c2} (4 ^3P_{2})$ &                     & 4613 &      & & & &\\
  $\chi_{c1} (4 ^3P_{1})$ &                     & 4605 &      & & & &\\
  $\chi_{c0} (4 ^3P_{0})$ &                     & 4586 &      & & & & \\
  $h_{c}(4 ^1P_{1})$    &                     & 4597 &      & & & & \\
  \hline\hline
\end{tabular}
\label{qspectrum}
\end{table}

\begin{table}[H]
\caption{Values of quenched and unquenched
 quark model parameters.}
\footnotesize
\centering
\begin{tabular}{lccccccccc}
  \hline\hline
   & $m_{c}$\;(GeV)\;\;  &\;\;  b\;(GeV$^{2}$)\;\; & \;\;$\mathcal{C}$ (GeV)\;\; & \;\;\; $\alpha_0$ \;\;\; & \;\;\; $\sigma_h$\;(GeV)\;\;\; &\;\;\; $\sigma_{S}$ (GeV) \;\;\;  \\
   \hline
   Quenched                      & 1.549 & 0.1436 & -0.0872 & 0.619 & 1.1502 & 0.2590  \\
 \hline
   \textbf{bare $\beta$}(1S)      & 1.729 & 0.1074 & 0.2611  & 1.222 & 0.9325 & 0.2255  \\
   \phantom{\textbf{bare $\beta$}}(2S)        & 1.998 & 0.0926 & 0.0160  & 1.103 & 1.1376 & 0.2598 \\
   \hline
   \textbf{consistent $\beta$}(1S)     & 1.435 & 0.0986 & 0.8145  & 1.183 & 0.7789 & 0.1914 \\
   \phantom{\textbf{consistent $\beta$}}(2S)       & 1.536 & 0.0913 & 0.7945  & 0.999 & 0.895  & 0.2055 \\
   \hline\hline
\end{tabular}
\label{parameters}
\end{table}

\subsection{Unquenching the Model}
\label{sec:unq}

\begin{figure}[H]
\caption{Loop diagram}
\emph{}
\label{loop}
\centering
\includegraphics[width=12 cm]{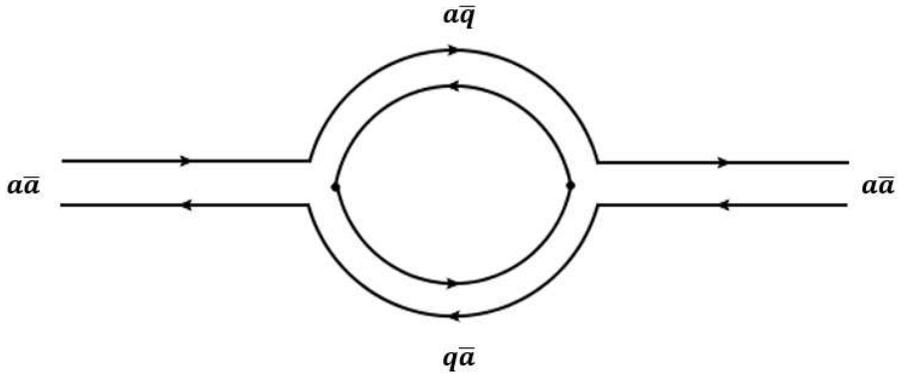}
\end{figure}

In the unquenched quark model, the process $A(a\overline{a})\rightarrow B(a\overline{q})+C(q\overline{a})\rightarrow A(a\overline{a})$ induces a hadron loop via transitory creation of a quark-antiquark pair ($q\bar{q}$) as shown in the Figure~\ref{loop}.

Loop effects are incorporated in the formalism with the $^{3}P_{0}$ model, first proposed by  Micu~\cite{micu} in 1969 and developed Le Yaouanc \textit{et al.} over many years~\cite{orsay}.
This model has been applied to study hadronic strong decays by many groups~\cite{decays} and to calculate hadron loop effects~\cite{loops}.

According to this model, a hadronic strong decay proceeds through the production of a quark-antiquark pair with vacuum quantum numbers ${}^{2S+1}L_J = {}^3P_0$. A compact way to write the operator is

\begin{equation}\label{3p0h}
    \widehat{H}_{I}= 2 m_{u/d}\gamma \int d^{3}x \overline{\psi}_{q}\psi_{q},
\end{equation}
where $m_{u/d}$ is the mass of the constituent up (or down) quark. Notice that this choice introduces a factor of $m_{u/d}/m_s$ when strange quark pairs are produced, which follows a suggestion of Kalashnikova\cite{K1}. The parameter $\gamma=0.35$, representing the amplitude of producing quark-antiquark pair from the vacuum, was determined in Ref.~\cite{hcharmonia} by  fitting the experimentally known hadronic decays of charmonium states.

We define the matrix element for a given strong decay,
$A\rightarrow B+C$, via

\begin{equation}\label{interaction matrix element}
    \langle BC|\hat{H}_{I}|A\rangle\;=\;h_{fi}\; \delta(\mathbf{P_A}-\mathbf{P_B}-\mathbf{P_C}),
\end{equation}
where $\mathbf{P_A},\mathbf{P_B}$ and $\mathbf{P_C}$ are the momenta of mesons $A$, $B$, and $C$ respectively.
With this definition the perturbative shift in the pole location for meson $A$ is given by\cite{08hloop}


\begin{multline}
\label{deltam}
  \Delta M^{BC}_{A}\;=\;\mathcal{P}\int^{\infty}_{M_{B}+M_{C}}\; \frac{dE_{BC}}{M_{A}-E_{BC}} \frac{P E_{B}E_{C}}{E_{BC}}\int d\Omega_{P}|h_{fi}|^{2}
  + i \pi \frac{P E_{B}E_{C}}{M_{A}} \int d\Omega_{P}|h_{fi}|^{2},
\end{multline}
where $P = P_A = P_B$ in the centre of mass frame and $\mathcal{P}$ denotes the principal value.

Nonperturbative estimates of the mass shift can be made in the absence of final state interactions by summing iterated bubble diagrams. The result is a full propagator of the form

\begin{equation}
-i G(s) = \frac{1}{s-M^2 - \Sigma(s)}
\end{equation}
where $i\Sigma$ is the one particle irreducible self-energy of the meson in question. The propagator pole yields the meson mass shift and width.
Computing numerical pole positions with iterated loops for several cases reveals differences of order 10\% compared to the perturbative result of Eq. \ref{deltam}. Since this is a reasonably small effect, we simply use the perturbative mass shift in the remainder of this work.
With this scheme, the mass of a conventional meson is given by

\begin{equation}\label{cs1}
    M_A=M_A^{(0)} + \sum_{B,C}\Re \Delta M^{BC}_{A},
\end{equation}
where $M_{A}^{(0)}$ is the quenched (``bare") meson mass and the sum runs over all meson pairs that couple to meson $A$.

The latter point is problematic because the sum can, in principle, diverge. In fact, the sum is logarithmically divergent when evaluated with simple harmonic oscillator (SHO) wavefunctions and the $^{3}P_{0}$ vertex of Eq. \ref{3p0h} \cite{GI1}. Of course, it is possible that this divergence can be absorbed in the model parameters. Nevertheless, the numerical value of the resulting mass shift is essentially unknown unless the sum converges very rapidly.
As far as we are aware, this issue has been ignored in the literature. One way forward is to regulate the decay amplitude by including a form factor in Eq. \ref{3p0h}, as is done, for example, in Refs. \cite{GI1, 17naeem, 18segovia}. In this case the form factor scale was treated as an observable and was fit to decay data.
Another approach would be to regulate the sum and remove the regulator dependence via renormalization. As far as we are aware, this has not been attempted.

Previous work in unquenching the quark model typically assumes that the sum in Eq. \ref{cs1} converges so rapidly that the first (or at most, a few) terms can be considered.
Even this must be done with some care because the sum over channels progresses in an uneven fashion. Specifically, Geiger and Isgur noticed the sum over a spin multiplet is independent of the spin quantum number of the decaying meson in the static (all quark masses large) limit for the $^{3}P_{0}$ model\cite{91GI}. This observation has been generalized to decay vertices with spin one that are factorisable--namely, a sum over a degenerate spin multiplet yields the same mass shift for all states, $A$, in a given $nL$ multiplet\cite{08hloop}.  This means that it is misleading to evaluate a mass shift due to a single hadron loop, such as $D\bar{D}$. Rather the full ground state multiplet, $(D\bar{D}$, $D^*\bar{D}$, $D^*\bar{D}^*$), should be considered. Another important implication is that the majority of mass shifts can be absorbed by the constant term in the model Hamiltonian, with only residual shifts due to spin splittings remaining.

Here we follow tradition in the field and simply truncate the sum. However we will sum over all members of a spin multiplet to account for the decay theorem mentioned above and will examine the effect of moving higher in radial quantum numbers. Furthermore, unlike in previous work, we incorporate effects due to the hadron loops into the wavefunction of the decaying meson, thereby making the model consistent.

Our calculation will be done with analytic expressions for the $^{3}P_{0}$ matrix elements obtained with simple harmonic oscillators. This choice simplifies the fit procedure, making the computation feasible. This approximation is reliable because only bulk behaviour of the wavefunctions is probed by the transition operator. The scale of the SHO wavefunctions, denoted $\beta$, and the relevant meson masses are reported in Table \ref{betacq}.

\begin{table}[H]
\caption{The SHO parameter $\beta$ and mass for $c\bar{q}$ meson states.}
\footnotesize
\centering
\begin{tabular}{c|ccccc}
  \hline\hline
  Meson state            & Mass (MeV) & Mass used in calculations (MeV)& $\beta$ (GeV)\cite{ishrat2} \\
  \hline
  $D$ $(1^1S_{0})$       & $D^{\pm}$ =1869.58$\pm$0.09 \cite{pdg}  & 1867   & 0.442\\
                              & $D^{0}$  =1864.83$\pm$0.08 \cite{pdg}   &        & \\
  $D^*$ $(1^3S_{1})$       & $D^{*\pm}$=2010.26$\pm$0.05\cite{pdg}   & 2008   & 0.338\\
                              & $D^{*0}$  =2006.85$\pm$0.05 \cite{pdg}  &        &\\
  $D$ $(2^1S_{0})$       & 2483                       \cite{15LY}  & 2483   & 0.328 \\
  $D^*$ $(2^3S_{1})$       & 2579.5 $\pm$ 3.4 $\pm$ 5.5\cite{13lhcb} & 2579.5 & 0.287 \\
  $D_{s}$ $(1^1S_{0})$   & $D_s^{\pm}$=1968.27$\pm$0.10 \cite{pdg}  & 1968   & 0.463\\
  $D^*_{s}$ $(1^3S_{1})$   & $D_s^{*\pm}$=2112.1$\pm$0.4  \cite{pdg}  & 2112   & 0.369\\
  $D_{s}$ $(2^1S_{0})$   & 2632.5 $\pm$ 1.7        \cite{04selex}  & 2632.5 & 0.348\\
  $D^*_{s}$ $(2^3S_{1})$   & 2708 $\pm$ $9_{10}^{+11}$\cite{08belle} & 2708   & 0.312\\
  \hline\hline
\end{tabular}
\label{betacq}
\end{table}

SHO wavefunctions will also be used for the charmonium states; in this case the SHO scales will be obtained by fitting to numerically computed solutions of the Schr\"{o}dinger equation. As mentioned above, past practice has been to fix these scales to the bare value, $\beta(m_{c}^{(0)},b^{(0)},\mathcal{C}^{(0)},\alpha_0^{(0)}, \sigma_h^{(0)}, \sigma_S^{(0)})$.  We will use this method in the following as well so that the results can be compared to the--more consistent--use of
 $\beta(m_{c},b,\mathcal{C},\alpha_0, \sigma_h, \sigma_S)$ as obtained in the fit. Convergence of the sum will be examined by truncating at the six lowest $nL = 1S$  $D$ and $D_s$ states and comparing the results to a sum over the twenty $D$ and $D_s$ combinations available up to $nL = 2S$.

\section{Results}

Our primary results are presented in Tables \ref{unqspectrumcs1} and \ref{unqspectrumcs2}. The tables reveal that mass shifts do indeed increase as the number of intermediate channels is increased, with mass shifts of about 8\% for 1S channels and 15\% for summing up to 2S. Nevertheless, the renormalized meson masses remain roughly constant in both cases. We find that this shift is largely absorbed by the quark mass when the bare SHO parameter is used, and by the Hamiltonian constant when the consistent SHO parameter is used.

A summary of the spectra for the unquenched calculation and the bare and consistent SHO parameter cases summing to 1S and 2S levels is given in Table \ref{allspectrum}. For convenience, we also show results from two other unquenching calculations in the final columns. The average deviation from experiment ranges from 12 to 33 MeV and is therefore quite good. It is apparent, however, that unquenching the model yields a slight degradation in fit quality. Nevertheless, it is notable that employing the consistent prescription for SHO parameter leads to a 10\% improvement in fit quality.   It is also reassuring that the average error decreases by approximately 20\% in moving from 1S (6 continuum) to 2S (20 continuum) channels, giving some indication that the method may be converging, albeit, slowly, and that it may ultimately improve the accuracy of the predictions.

We remark that most of the predicted charmonia masses are relatively stable on moving from the 1S to 2S sum over intermediate states, with most renormalization effects occurring higher in the spectrum. Interestingly, the $\chi_{cJ}(2P)$ multiplet is shifted upward by approximately 40 MeV in going from the 1S to 2S sum. This sensitivity is doubtlessly due to the proximity of the $D^*\bar{D}$ threshold, and clearly plays an important role in properties of the $X(3872)$.

The $\chi_{cJ}(nP)$ multiplets are split by tensor and spin-orbit interactions in the quark model presented here (and most others). Experimentally, the weighted center of mass of the multiplet lies very close to the partner $h_c(nP)$ mass, namely the quantity

\begin{equation} \label{eq:DeltaDef}
\Delta \equiv M_h - \frac{1}{9} \left[ 1 \! \cdot \! M_{\chi_0} + 3 \! \cdot \! M_{\chi_1} + 5 \!
    \cdot \! M_{\chi_2} \right] \, ,
\end{equation}
is very small\cite{ls}. Indeed, $\Delta$ is an astonishing 0.08 MeV for the 1P charmonium multiplet. The quenched quark model used here yields $\Delta(1P)$ = -10 MeV. This relatively large splitting is due to the coordinate smearing used in the model, which shifts matrix elements between the states slightly.

The increased sensitivity due to the nearby $D^*\bar{D}$ may also be expected to cause shifts in spin splittings. As mentioned in Section \ref{sec:unq}, in general one expects loop effects to recapitulate continuum spin splittings in the renormalized meson masses. This is  important since it implies that the spin-dependent interaction already present in the model can absorb these effects. However, in general residual spin-dependent mass shifts can occur. We see that this is indeed the case for the $\chi_{cJ}$ multiplets. In particular, the values of $\Delta$ get somewhat larger upon unquenching, as shown in Table \ref{tab:delta}. Notice that the values of $\Delta$ appear to be slowly moving towards the bare values as the number of intermediate states is increased, which is again an indication of the slow convergence of the formalism, and perhaps, its stability with respect to the bare model.

\begin{table}[h]
\caption{Ultrafine splittings for $\chi_{cJ}(nP)$ multiplets for the consistent SHO parameter case (MeV).}
\label{tab:delta}
\begin{tabular}{c|cccc}
\hline\hline
$n$ & expt & bare & $\Delta(1S)$ & $\Delta(2S)$ \\
\hline
1 & 0.08 & -10 & -25 & -21 \\
2 &      & -11 & -21 & -18\\
\hline\hline
\end{tabular}
\end{table}

Lastly, the $\chi_{c2}(1P)$ state is the heaviest in its multiplet, and the bare quark model indicates that this should remain true for the radially excited multiplets. We find that this pattern of split splittings remains in the unquenched model, \textit{except} for the 2P case, where the $\chi_{c1}(2P)$ mass shifts higher than the $\chi_{c2}(2P)$. Again, this is most likely due to the proximity of the $D^*\bar{D}$ channel.

\begin{table}[H]
\caption{Unquenched charmonium spectrum and mass shifts (in MeV). Computed with bare SHO parameters (last column). Light quark masses are $m_{u/d}$=$0.33$ GeV and $m_{s}$=$0.55$ GeV.}
\centering
\footnotesize
\begin{tabular}{cc|ccc|ccc|c}
\hline\hline
Meson state& Exp. Mass  &  $M$  & $M_{0}$ & $\Delta M(1S)$ &  $M$  & $M_{0}$ & $\Delta M(2S)$ & $\beta_A^{(0)}$(GeV) \\
\hline
   $J/\psi (1 ^3 S_{1}) $     & 3096.9 $\pm$ 0.006 & 3089& 3310  & -221.0 & 3090 & 3552 & -462.5 & 0.602\\
   $\eta_{c}(1 ^1S_{0})$     & 2983.9 $\pm$ 0.5   & 2979& 3186  & -207.2 & 2980 & 3420 & -439.9 & 0.682\\
   $\psi' (2 ^3S_{1})$        & 3686.097 $\pm$ 0.025&3679& 3976  & -297.2 & 3682 & 4191 & -509.0 & 0.497\\
   $\eta^{'}_{c}(2 ^1S_{0})$  & 3637.6 $\pm$ 1.2   & 3659& 3943  & -283.5 & 3651 & 4157 & -505.5 & 0.529\\
   $\psi$ $(3^3S_{1})$       & 4039 $\pm$ 1       & 4070& 4366  & -295.6 & 4092 & 4541 & -449.3 & 0.435\\
   $\psi$ $(4^3S_{1})$       & 4421 $\pm$ 4       & 4471& 4668  & -196.8 & 4435 & 4807 & -372.6 & 0.4 \\
   $\chi_{c2} (1 ^3P_{2})$    & 3556.17 $\pm$ 0.07 & 3552& 3835  & -282.8 & 3553 & 4080 & -526.7 & 0.467\\
   $\chi_{c1} (1 ^3P_{1})$   & 3510.67 $\pm$ 0.05 & 3530& 3793  & -263.2 & 3527 & 4031 & -503.8 & 0.479\\
   $\chi_{c0} (1 ^3P_{0})$   & 3414.71 $\pm$ 0.3  & 3468& 3728  & -259.4 & 3456 & 3961 & -505.1 & 0.498\\
   $h_{c}(1 ^1P_{1})$        & 3525.38 $\pm$ 0.1  & 3516& 3784  & -268.2 & 3520 & 4032 & -512.0 & 0.484\\
   $\psi (1 ^3D_{1})$       & 3773.13 $\pm$ 0.35   & 3706& 4042 & -336.3 & 3708 & 4258 & -550.1 & 0.449\\
   $\psi (2 ^3D_{1})$       & 4191 $\pm$ 5         & 4114& 4409 & -295.1 & 4132 & 4586 & -454.4 & 0.418\\
   $\eta_{c}(3 ^1S_{0})$     &                    & 4088& 4346  & -257.8 & 4105 & 4521 & -416.1 & 0.451 \\
   $\eta_{c}(4 ^1S_{0})$      &                    & 4463& 4652  & -190.0 & 4420 & 4792 & -371.5 & 0.409\\
   $\psi$ $(5^3S_{1})$        &                    & 4732& 4926  & -194.0 & 4677 & 5034 & -356.8 & 0.376\\
   $\eta_{c}(5 ^1S_{0})$      &                    & 4723& 4913  & -190.6 & 4683 & 5021 & -338.3 & 0.383\\
   $\chi_{c2} (2 ^3P_{2})$    &                    & 3927& 4255  & -328.1 & 3957 & 4451 & -526.7 & 0.429\\
   $\chi_{c1} (2 ^3P_{1})$   &                    & 3984& 4231  & -246.9 & 4015 & 4426 & -411.2 & 0.437\\
   $\chi_{c0} (2 ^3P_{0})$   &                    & 3903& 4193  & -289.5 & 3928 & 4389 & -460.2 & 0.451\\
   $h_{c}(2 ^1P_{1})$        &                    & 3929& 4223  & -293.6 & 3964 & 4422 & -460.2 & 0.441\\
   $\chi_{c2} (3 ^3P_{2})$    &                    & 4363& 4568  & -204.8 & 4352 & 4726 & -374.0 & 0.397\\
   $\chi_{c1} (3^3P_{1})$    &                    & 4773& 4554  & -219.2 & 4330 & 4711 & -380.8 & 0.403\\
   $\chi_{c0} (3 ^3P_{0})$   &                    & 4335& 4529  & -194.6 & 4325 & 4688 & -362.6 & 0.412\\
   $h_{c}(3 ^1P_{1})$        &                    & 4339& 4545  & -205.8 & 4338 & 4706 & -367.5 & 0.406\\
\hline\hline
\end{tabular}
\label{unqspectrumcs1}
\end{table}

\begin{table}[H]
\footnotesize
Table~\ref{unqspectrumcs1}\vspace{0.5cm}~Continued.
\newline
\centering
\begin{tabular}{cc|ccc|ccc|c}
\hline\hline
Meson state& Exp. Mass   &  $M$  & $M_{0}$ & $\Delta M(1S)$ &  $M$  & $M_{0}$ & $\Delta M(2S)$ & $\beta_A^{(0)}$(GeV) \\
\hline
   $\chi_{c2} (4 ^3P_{2})$    &                    & 4634& 4834  & -200.0 & 4612 & 4959 & -346.4 & 0.375\\
   $\chi_{c1} (4 ^3P_{1})$   &                    & 4629& 4824  & -195.4 & 4559 & 4948 & -388.5 & 0.379\\
   $\chi_{c0} (4 ^3P_{0})$   &                    & 4608& 4806  & -198.2 & 4509 & 4930 & -420.7 & 0.385\\
   $h_{c}(4 ^1P_{1})$        &                    & 4621& 4815  & -194.5 & 4578 & 4942 & -364.1 & 0.381\\
   $\chi_{c2} (5 ^3P_{2})$    &                    & 4881& 5071  & -190.4 & 4837 & 5165 & -328.7 & 0.359\\
   $\chi_{c1} (5 ^3P_{1})$   &                    & 4889& 5064  & -175.1 & 4837 & 5156 & -319.1 & 0.362\\
   $\chi_{c0} (5 ^3P_{0})$   &                    & 4869& 5049  & -180.5 & 4832 & 5142 & -310.5 & 0.367\\
   $h_{c}(5 ^1P_{1})$        &                    & 4877& 5056  & -178.8 & 4829 & 5151 & -321.8 & 0.364\\
   $\psi_{3} (1 ^3D_{3})$  &                      & 3850& 4157 & -307.5 & 3883 & 4376 & -492.3 & 0.410 \\
   $\psi_{2} (1 ^3D_{2})$   &                      & 3809& 4105 & -295.8 & 3830 & 4322 & -491.6 & 0.428\\
   $\eta_{c2}(1 ^1D_{2})$    &                      & 3819& 4114 & -295.0 & 3846 & 4334 & -488.0 & 0.425\\
   $\psi_{3} (2 ^3D_{3})$    &                      & 4220& 4477 & -256.2 & 4287 & 4651 & -364.5 & 0.392\\
   $\psi_{2} (2 ^3D_{2})$   &                      & 4190& 4446 & -256.2 & 4215 & 4622 & -406.6 & 0.404\\
   $\eta_{c2}(2 ^1D_{2})$    &                      & 4219& 4449 & -230.4 & 4245 & 4627 & -381.3 & 0.402\\
   $\psi_{3} (3 ^3D_{3})$  &                      & 4542& 4747 & -204.4 & 4491 & 4887 & -395.5 & 0.373\\
   $\psi_{2} (3 ^3D_{2})$   &                      & 4532& 4726 & -194.3 & 4477 & 4866 & -387.6 & 0.381\\
   $\psi (3 ^3D_{1})$       &                      & 4496& 4701 & -205.1 & 4469 & 4842 & -373.0 & 0.391\\
   $\eta_{c2}(3 ^1D_{2})$    &                      & 4534& 4727 & -192.3 & 4482 & 4869 & -386.3 & 0.380\\
   $\psi_{3} (4 ^3D_{3})$  &                      & 4794& 4988 & -194.3 & 4771 & 5097 & -325.5 & 0.358\\
   $\psi_{2} (4 ^3D_{2})$   &                      & 4801& 4973 & -171.4 & 4775 & 5080 & -305.5 & 0.364\\
   $\psi (4 ^3D_{1})$       &                      & 4781& 4954 & -172.2 & 4772 & 5061 & -289.1 & 0.371\\
   $\eta_{c2}(4 ^1D_{2})$    &                      & 4796& 4972 & -175.4 & 4767 & 5082 & -314.8 & 0.363\\
  \hline\hline
\end{tabular}
\end{table}

\begin{table}[H]
\caption{Unquenched charmonium spectrum and mass shifts (in MeV). Computed with consistent SHO parameters.  Light quark masses are $m_{u/d}$=$0.33$ GeV and $m_{s}$=$0.55$ GeV.}
\footnotesize
\centering
\begin{tabular}{cc|cccc|cccc}
  \hline\hline
      Meson state             & Exp. Mass  & $M$  & $M_{0}$ & $\Delta M(1S)$ & $\beta_{A}$ (GeV) &  $M$ & $M_{0}$ & $\Delta M(2S)$ & $\beta_{A}$(GeV) \\
  \hline
   $J/\psi (1 ^3 S_{1}) $     & 3096.9 $\pm$ 0.006    & 3093 & 3347 & -253.9 & 0.619 & 3096 & 3570 & -474.4 & 0.638\\
   $\eta_{c}(1 ^1S_{0})$      & 2983.9 $\pm$ 0.5      & 2981 & 3219 & -238.6 & 0.687 & 2985 & 3437 & -452.8 & 0.723 \\
   $\psi' (2 ^3S_{1})$        & 3686.097 $\pm$ 0.025  & 3666 & 3994 & -328.1 & 0.491 & 3663 & 4176 & -512.3 & 0.495\\
   $\eta^{'}_{c}(2 ^1S_{0})$  & 3637.6 $\pm$ 1.2      & 3648 & 3961 & -313.3 & 0.517 & 3634 & 4142 & -509.2 & 0.526 \\
   $\psi$ $(3^3S_{1})$       & 4039 $\pm$ 1          & 4066 & 4373 & -306.9 & 0.415 & 4090 & 4523 & -433.4 & 0.415 \\
   $\psi$ $(4^3S_{1})$       & 4421 $\pm$ 4          & 4447 & 4670 & -223.2 & 0.372 & 4421 & 4796 & -375.0 & 0.371\\
   $\chi_{c2} (1 ^3P_{2})$    & 3556.17 $\pm$ 0.07    & 3551 & 3865 & -313.8 & 0.466 & 3551 & 4076 & -525.1 & 0.458 \\
   $\chi_{c1} (1 ^3P_{1})$   & 3510.67 $\pm$ 0.05    & 3534 & 3825 & -291.0 & 0.472 & 3531 & 4032 & -501.7 & 0.467\\
   $\chi_{c0} (1 ^3P_{0})$   & 3414.71 $\pm$ 0.3     & 3477 & 3762 & -284.7 & 0.482 & 3464 & 3965 & -501.2 & 0.481\\
   $h_{c}(1 ^1P_{1})$        & 3525.38 $\pm$ 0.1     & 3512 & 3813 & -300.8 & 0.492 & 3514 & 4030 & -516.1 & 0.482 \\
   $\psi (1 ^3D_{1})$       & 3773.13 $\pm$ 0.35     & 3709 & 4062 & -353.5 & 0.421 & 3718 & 4242 & -524.6 & 0.419 \\
   $\psi (2 ^3D_{1})$       & 4191 $\pm$ 5           & 4136 & 4418 & -282.1 & 0.389 & 4157 & 4570 & -412.7 & 0.387 \\
   $\eta_{c}(3 ^1S_{0})$     &              & 4080 & 4353 & -273.5 & 0.427 & 4098 & 4503 & -404.0 & 0.429 \\
   $\eta_{c}(4 ^1S_{0})$      &               & 4437 & 4655 & -218.0 & 0.379 & 4404 & 4781 & -376.8 & 0.379 \\
   $\psi$ $(5^3S_{1})$        &               & 4718 & 4926 & -207.7 & 0.345 & 4693 & 5032 & -338.7 & 0.344\\
   $\eta_{c}(5 ^1S_{0})$      &               & 4712 & 4913 & -201.4 & 0.350 & 4683 & 5019 & -335.7 & 0.349\\
   $\chi_{c2} (2 ^3P_{2})$    &               & 3928 & 4269 & -340.2 & 0.466 & 3972 & 4437 & -465.4 & 0.405\\
   $\chi_{c1} (2 ^3P_{1})$   &               & 3979 & 4246 & -266.4 & 0.418 & 4017 & 4413 & -396.6 & 0.414\\
   $\chi_{c0} (2 ^3P_{0})$   &               & 3897 & 4207 & -310.3 & 0.428 & 3927 & 4375 & -447.3 & 0.427 \\
   $h_{c}(2 ^1P_{1})$        &               & 3921 & 4236 & -314.3 & 0.429 & 3964 & 4408 & -443.6 & 0.422\\
   $\chi_{c2} (3 ^3P_{2})$    &               & 4343 & 4575 & -231.3 & 0.371 & 4336 & 4716 & -379.4 & 0.366\\
   $\chi_{c1} (3^3P_{1})$    &               & 4320 & 4561 & -240.8 & 0.376 & 4320 & 4701 & -381.1 & 0.372\\
   $\chi_{c0} (3 ^3P_{0})$   &               & 4314 & 4536 & -222.5 & 0.384 & 4305 & 4677 & -371.7 & 0.382\\
   $h_{c}(3 ^1P_{1})$        &               & 4322 & 4551 & -229.0 & 0.382 & 4324 & 4694 & -370.0 & 0.377\\
   \hline\hline
\end{tabular}
\label{unqspectrumcs2}
\end{table}

\begin{table}[H]
\footnotesize
Table~\ref{unqspectrumcs2}\vspace{0.5cm}~Continued.
\newline
\centering
\begin{tabular}{cc|cccc|cccc}
  \hline\hline
      Meson state             & Exp. Mass  & $M$  & $M_{0}$ & $\Delta M(1S)$ & $\beta_{A}$ (GeV) &  $M$ & $M_{0}$ & $\Delta M(2S)$ & $\beta_{A}$(GeV) \\
  \hline
   $\chi_{c2} (4 ^3P_{2})$    &           & 4618 & 4837 & -218.5 & 0.466 & 4607 & 4956 & -348.9 & 0.341 \\
   $\chi_{c1} (4 ^3P_{1})$   &            & 4631 & 4827 & -195.9 & 0.348 & 4585 & 4945 & -360.1 & 0.345 \\
   $\chi_{c0} (4 ^3P_{0})$   &            & 4606 & 4809 & -202.9 & 0.354 & 4523 & 4927 & -404.1 & 0.352 \\
   $h_{c}(4 ^1P_{1})$        &            & 4615 & 4818 & -203.2 & 0.352 & 4592 & 4938 & -346.2 & 0.348  \\
   $\chi_{c2} (5 ^3P_{2})$    &           & 4833 & 5072 & -238.8 & 0.326 & 4801 & 5172 & -370.4 & 0.323 \\
   $\chi_{c1} (5 ^3P_{1})$   &            & 4850 & 5064 & -214.5 & 0.328 & 4812 & 5164 & -352.0 & 0.326 \\
   $\chi_{c0} (5 ^3P_{0})$   &            & 4847 & 5050 & -202.6 & 0.333 & 4826 & 5149 & -323.1 & 0.331\\
   $h_{c}(5 ^1P_{1})$        &            & 4841 & 5056 & -214.4 & 0.331 & 4811 & 5157 & -345.5 & 0.328\\
   $\psi_{3} (1 ^3D_{3})$  &            & 3852 & 4175 & -323.0 & 0.393 & 3899 & 4358 & -459.0 & 0.383 \\
   $\psi_{2} (1 ^3D_{2})$   &           & 3812 & 4124 & -311.5 & 0.407 & 3842 & 4305 & -463.3 & 0.401 \\
   $\eta_{c2}(1 ^1D_{2})$    &          & 3820 & 4132 & -312.2 & 0.407 & 3856 & 4316 & -460.5 & 0.398 \\
   $\psi_{3} (2 ^3D_{3})$    &          & 4248 & 4484 & -237.1 & 0.366 & 4261 & 4636 & -375.0 & 0.359 \\
   $\psi_{2} (2 ^3D_{2})$   &           & 4197 & 4455 & -258.3 & 0.377 & 4224 & 4606 & -382.3 & 0.372 \\
   $\eta_{c2}(2 ^1D_{2})$    &          & 4218 & 4457 & -239.2 & 0.377 & 4246 & 4611 & -365.0 & 0.371 \\
   $\psi_{3} (3 ^3D_{3})$  &            & 4523 & 4750 & -227.4 & 0.342 & 4491 & 4879 & -387.6 & 0.337 \\
   $\psi_{2} (3 ^3D_{2})$   &           & 4537 & 4731 & -193.7 & 0.350 & 4502 & 4858 & -355.9 & 0.346 \\
   $\psi (3 ^3D_{1})$       &           & 4510 & 4705 & -196.1 & 0.359 & 4483 & 4833 & -350.4 & 0.356 \\
   $\eta_{c2}(3 ^1D_{2})$    &          & 4529 & 4730 & -200.7 & 0.350 & 4494 & 4860 & -365.7 & 0.345 \\
   $\psi_{3} (4 ^3D_{3})$  &            & 4754 & 4989 & -235.2 & 0.324 & 4738 & 5098 & -359.4 & 0.320 \\
   $\psi_{2} (4 ^3D_{2})$   &           & 4771 & 4974 & -203.2 & 0.330 & 4739 & 5082 & -342.5 & 0.327 \\
   $\psi (4 ^3D_{1})$       &           & 4774 & 4955 & -181.2 & 0.337 & 4762 & 5063 & -301.0 & 0.334 \\
   $\eta_{c2}(4 ^1D_{2})$    &          & 4763 & 4972 & -209.5 & 0.331 & 4742 & 5082 & -340.3 & 0.327 \\
  \hline\hline
\end{tabular}
\end{table}

\begin{table}[H]
\footnotesize
\caption{The observed, quenched and unquenched charmonium spectrum (MeV). Columns 8 and 9 present masses computed in other unquenched models for comparison.}
\centering
\begin{tabular}{cc|c|cc|cc|cc}
  \hline\hline
\multirow{2}{*}{  } & \multirow{2}{*}{  } & \multicolumn{1}{r|}{ Quenched } &\multicolumn{2}{r|}{ \textbf{bare $\beta$} \;\;\;}&\multicolumn{2}{r|}{ \textbf{consistent $\beta$}  }& \multirow{2}{*}{  }\\
      Meson state             & Exp. Mass~\cite{pdg} & $M$ & $M(1S)$ & $M(2S)$ &$M(1S)$ & $M(2S)$ & Ref.~\cite{09BCK} & Ref.~\cite{13feretti}  \\
  \hline
  $J/\psi (1 ^3 S_{1}) $ & 3096.9 $\pm$ 0.006   & 3090 & 3089 & 3090 & 3093 & 3096 & 3100 & 3137 \\
   $\eta_{c}(1 ^1S_{0})$  & 2983.9 $\pm$ 0.5    & 2983 & 2979 & 2980 & 2981 & 2985 & 2980 & 2979 \\
   $\psi' (2 ^3S_{1})$    & 3686.097 $\pm$ 0.025& 3688 & 3679 & 3682 & 3666 & 3663 & 3674 & 3640 \\
 $\eta^{'}_{c}(2 ^1S_{0})$& 3637.6 $\pm$ 1.2    & 3648 & 3659 & 3651 & 3648 & 3633 & 3635 & 3588 \\
    $\psi$ $(3^3S_{1})$   & 4039 $\pm$ 1        & 4091 & 4070 & 4092 & 4066 & 4090 &  & \\
    $\psi$ $(4^3S_{1})$   & 4421 $\pm$ 4        & 4426 & 4471 & 4435 & 4447 & 4421 &  & \\
   $\chi_{c2} (1 ^3P_{2})$& 3556.17 $\pm$ 0.07  & 3555 & 3552 & 3553 & 3551 & 3551 & 3565 & 3527\\
   $\chi_{c1} (1 ^3P_{1})$& 3510.67 $\pm$ 0.05  & 3513 & 3530 & 3527 & 3534 & 3531 & 3520 & 3494\\
   $\chi_{c0} (1 ^3P_{0})$& 3414.71 $\pm$ 0.3   & 3442 & 3468 & 3456 & 3477 & 3464 & 3441 & 3430 \\
    $h_{c}(1 ^1P_{1})$    & 3525.38 $\pm$ 0.11  & 3518 & 3516 & 3520 & 3512 & 3514 & 3531 & 3501 \\
       $\psi (1 ^3D_{1})$ & 3773.13$\pm$0.35    & 3743 & 3706 & 3708 & 3709 & 3718 & 3794 & 3750\\
        $\psi (2 ^3D_{1})$& 4191 $\pm$ 5        & 4135 & 4114 & 4132 & 4136 & 4157 &  & \\
   \hline
   average error (MeV)    &                     & 17 & 29 & 24 & 26 & 21 & 12 & 26 \\
       \hline\hline
\end{tabular}
\label{allspectrum}
\end{table}

\begin{figure}[H]
\caption{Quark model parameters vs. $\gamma$ for the consistent SHO parameter case with 1S intermediate channels.}
\label{gammaplot}
\centering
\includegraphics[width=14 cm]{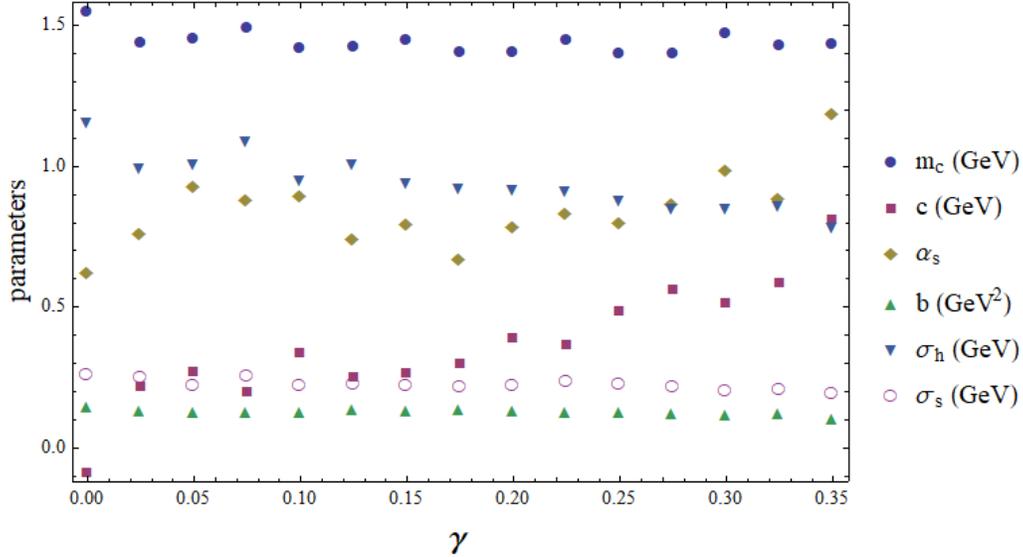}
\end{figure}

As we have mentioned, the bulk of the effects due to unquenching the quark model can be subsumed into the model parameters, and it is of interest to examine the way in which this occurs. We have done this by tracking the model parameters as a function of the strength of the quark pair creation vertex, $\gamma$. Figure \ref{gammaplot} shows the results. These were obtained by starting from the bare computation and stepping $\gamma$ up by small amounts, while refitting at each step. Thus a smooth transition from the bare to the unquenched model should be traced. The figure shows a strong general trend (modulo fluctuations due to the complexities of the objective function in the fit) upwards for the Hamiltonian constant term. Of course, this is anticipated. Interestingly, the string tension ($b$), smearing parameter ($\sigma_S$), and quark mass ($m_c$) remain largely constant. Alternatively, the strong coupling ($\alpha_s$) has an upward shift, while the hyperfine smearing ($\sigma_h$) has a slight downward trajectory. These effects tend to cancel in the hyperfine term, indicating that hyperfine splitting systematics may be preserved under unquenching, which is of course desirable.

\section{Conclusions}

We have computed the spectrum of low-lying charmonium states accounting for mixing with
 the open charm continuum with a simple and  phenomenologically successful model of strong decays.
We have noted that the sum over intermediate states should be organized by spin multiplet to leverage the ``cancellation theorem" of Ref. \cite{08hloop}. We have confirmed that summing over more continuum channels (than the usual lowest multiplet) leads to larger bare masses, but brings the spectrum closer to experiment. We have also shown that using the shifted charmonium wavefunction (expressed via a consistent SHO scale) in the coupling matrix element  improves the scheme.

The bulk of the induced mass shifts can be absorbed into the parameters of the model. The renormalized meson masses exhibit residual shifts that reflect the effects due spin splittings in the open charm mesons and the proximity of thresholds. In most cases mass shifts are small and have little bearing on the interpretation of charmonia. However in some cases, like that of $\chi_{cJ}(2P)$ multiplets, this is not true. These multiplets are important for the presence of the nearby and exotic $X(3872)$, which is often interpreted as a $D^{0*}\bar{D}^0$ bound state with a large $c\bar{c}$ component. A novel prediction of this work is that the related $\chi_{c1}(2P)$ (mostly) charmonium state should be higher in mass (by 150 MeV or more), and should be higher in mass than its partner $\chi_{c2}(2P)$ state.

Of course coupling  to the continuum does more than shift quark model mass predictions. Charmonium properties--especially for those near thresholds--should also change. These include lineshapes, decay strengths, electroweak transitions, and lifetimes. For example, such couplings could contribute an $s\bar{s}$ component to nuclear observables\cite{Bijker:2012zza}. They could also improve the description of hadronic transitions in bottomonia\cite{Zhou}.

Future work should address the, as yet unresolved, technical issues concerning the sum over virtual states and renormalization of the full quark model. Once this is achieved a consistent comparison to observables such as pole locations and strong and electroweak couplings can be made.

\textbf{Acknowledgment:} SK and FA would like to thank M.A. Sultan for providing assitance in developing a computer program at the early stage of the work. FA and BM are thankful for PU research grants 2021-22.

\end{document}